# EXPERIENCE WITH LONG-PULSE OPERATION OF THE PIP2IT WARM FRONT END**

A. Shemyakin†, J.-P. Carneiro, A. Chen, D. Frolov, B. Hanna, R. Neswold, L. Prost, G. Saewert, A. Saini, V. Scarpine, A. Warner, J.-Y. Wu, Fermilab, Batavia, IL 60510, USA
C. Richard, Michigan State University, East Lansing, MI, USA

## Abstract

The warm front end of the PIP2IT accelerator, assembled and commissioned at Fermilab, consists of a 15 mA DC, 30 keV H$^-$ ion source, a 2 m long Low Energy Beam Transport (LEBT) line, and a 2.1 MeV, 162.5 MHz CW RFQ, followed by a 10 m long Medium Energy Beam Transport (MEBT) line. A part of the commissioning efforts involves operation with the average beam power emulating the operation of the proposed PIP-II accelerator, which will have a duty factor of 1.1% or above. The maximum achieved power is 5 kW (2.1 MeV x 5 mA x 25 ms x 20 Hz). This paper describes the difficulties encountered and some of the solutions that were implemented.

## INTRODUCTION

The central part of Fermilab's Proton Improvement Plan, Stage Two (PIP-II) project [1] is an 800 MeV, 2 mA CW-compatible superconducting H$^-$ linac. A prototype of the PIP-II linac front end called PIP-II Injector Test (PIP2IT) is being built to mitigate technical risks associated with acceleration at low energies and to demonstrate a capability to create an arbitrary bunch structure. The PIP2IT Warm Front End (WFE, Fig. 1) comprised of the LEBT, RFQ, and MEBT followed by a 20 kW beam dump has been assembled and commissioned [2]. In 2019, two cryomodules will be installed downstream.

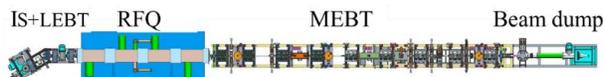

Figure 1: PIP2IT warm front end (top view).

While most of the beam properties were measured with a short pulse length (~10 μs) to minimize the beam's damage potential, stable operation with parameters required for the initial mode of operation of the PIP-II linac with Booster injection only (1.1% duty factor), was demonstrated during the Spring'18 run. In addition, since all major components of PIP-II are being designed to be CW-compatible, the duty factor was pushed significantly higher during a series of dedicated tests.

## HIGH DUTY-FACTOR MODE

### Ion source and LEBT

The ion source and LEBT have been commissioned to reliably deliver to the RFQ a 30 keV beam of up to 10 mA with pulse lengths of 1 μs - 50 ms at 20 Hz repetition rate defined by the LEBT chopper, as well as a true DC beam. An atypical LEBT transport scheme [3] minimizes changes of the beam properties throughout the pulse due to neutralization, which allows tuning of the downstream accelerator with a short pulse length of 10 μs or less. Reliability of the ion source [4] and LEBT did not significantly affect the overall performance of the machine in long-pulse operation. During the 3-month period of the Spring'18 run, the ion source experienced only one spark, which was a dramatic improvement (by a factor of ~100 in the rate) in comparison with the beginning of the ion source operation in 2013-2015.

### RFQ

While the RFQ has been designed for CW operation [5], it was used mainly in the pulse mode, partly as an additional protection from accidental generation of a long-pulse beam, and partly because of concerns about the power couplers since two out of four initially manufactured ceramic windows developed vacuum leaks after CW operation. In preparation for the Spring'18 run, the window design was modified to use O-ring sealing instead of brazing [6]. The newly designed couplers worked reliably, with the accumulated time of CW operation reaching 350 hours. The time to bring the RFQ to its nominal frequency from a cold state using the resonance control system [7] was ~20 min.

After conditioning the RFQ in CW to 65 kV of intervane voltage, the average rate of sparks at the nominal 60 kV was about once per hour, although this varied greatly from day to day. Following a spark, the RF power is recovered by the next beam pulse if the internal trip counter stays below 10. The RFQ then typically remains in resonance, and beam operation can resume quickly. If there are no sparks over a sliding one-minute window, the counter is reset to zero. If the counter reaches 10, the protection system turns the RFQ power off, and recovery requires the intervention of an operator. In such infrequent cases, the recovery time depends on how quickly the recovery starts and varies from minutes to tens of minutes.

One of the unexpected features was short bursts of the RFQ vacuum happening independently on the presence of RF. In the worst cases, the pressure in the RFQ gauges went above $10^{-6}$ Torr, and ~10% of the beam was lost (Fig.2). These jumps were eventually traced to excessive grease on the large O-rings of the RFQ vacuum flanges. Air permeating through the O-rings was accumulating in local bubbles created by the grease and released into the vacuum chamber in bursts. The bursts were mostly eliminated after removing the grease.

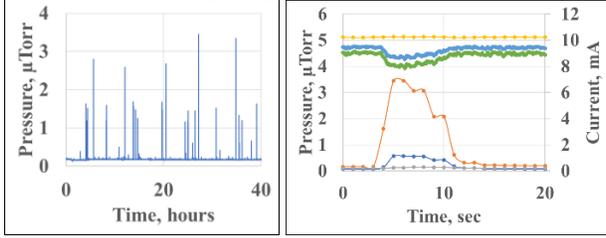

Figure 2: Vacuum bursts in the RFQ. Left- reading of the vacuum gauge in the 4[th] RFQ section during a 40-hour period. Right – detailed view of one the bursts with 10 mA beam running. Beam current is shown in LEBT (yellow), after the RFQ (blue), and in the dump (green). Pressure is shown in the 4[th] RFQ section (orange), 1[st] RFQ section (blue), and in the 1[st] bunching cavity (grey).

*MEBT*

The MEBT configuration in Spring'18 is shown in Fig. 3 and described in Ref. [8]. Difficulties with long-pulse operation in the MEBT are mainly associated with the presence of the chopping system [9] that removes pre-selected bunches. The kickers need to be protected from being damaged by the beam, while they set tight vertical aperture restrictions. In addition, to protect the future cryomodules from the high gas load generated by the bunches deflected onto the absorber, a Differential Pumping Insert (DPI) is installed just downstream of the absorber section. Thus, the DPI's 200 mm (L) × 10 mm (ID) beam pipe is another tight aperture restriction.

After tuning the beam envelope and adjusting the trajectory, a short-pulse beam was passed through the entire MEBT with low losses (below the beam loss monitoring system's resolution of ~4%) both with kickers off and while pulsing. The corresponding simulated beam envelope (Fig. 4) agrees well with the measurements [8].

For long-pulse operation, protection against beam-induced damage was provided through several layers, starting with administrative procedures and robust designs. The kickers were designed to withstand a heat load of 40 W, well above the heat generated by the electrical pulses in the version of the kicker finally chosen. Furthermore, while the gap between the kicker electrodes is 16 mm, the aperture is restricted by 13 mm-high slits in plates installed at both the entrance and exit of each kicker structure. These protection electrodes are electrically isolated, and their current is reported to the Machine Protection System (MPS) [11]. Similarly, the DPI can dissipate at least 25 W of power, also with the capability to report the loss current.

The next layer of protection uses the scraper system [12], which consists of 4 sets of 4 movable, electrically isolated, radiation-cooled 75 W-rated scrapers. Before switching to long-pulse mode, all 8 scrapers in the first two sets are positioned at the beam boundary, typically removing in total 1-2% of the beam. In addition, if the beam shifts or its envelope changes, the resulting increase of the scraper current induces a beam interruption via the MPS. Usually, downstream scrapers were placed at the beam boundary as well to emulate protection of future cryomodules downstream.

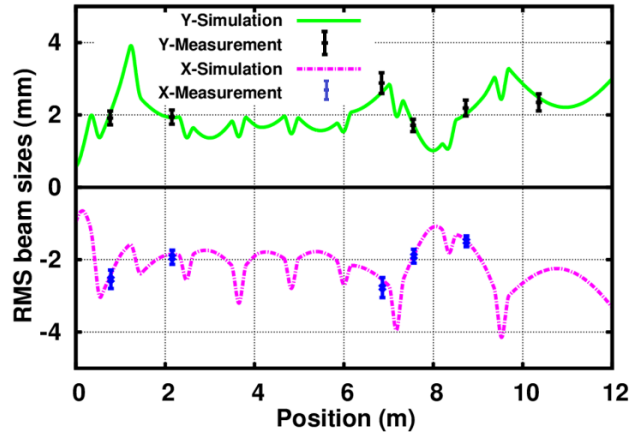

Figure 4: Rms beam envelope along the MEBT simulated with TRACEWIN [10]. Normalized rms transverse and longitudinal emittances are 0.2 µm and 0.28 µm, respectively. Beam current is 5 mA.

## MACHINE PROTECTION SYSTEM

The final layer of protection is provided by the MPS. When it receives a triggering signal from one of its channels, it interrupts the beam by dropping the ion source extraction voltage and raising the LEBT chopper voltage that prevents the beam from entering the RFQ. In the most severe cases (e.g. sensing vacuum deterioration), and for redundancy, the ion source high voltage and the LEBT bending magnet are turned off as well.

In the Spring'18 run, the MPS channels included signals from vacuum and water systems, readiness of the LEBT kicker, RF interlocks, MEBT kicker protection electrodes, DPI, and a beam current – comparing system. The latter is based on comparing readings from current-sensitive devices in various parts of the MEBT, ideologically similar to Ref. [13].

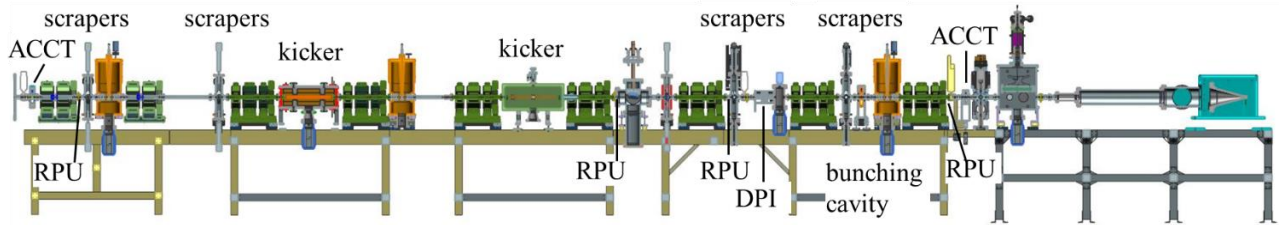

Figure 3: Medium Energy Beam Transport line (side view).

This system is based on electrostatic pickups, so-called Ring Pickups (RPU), with two pairs of RPUs monitoring the beam propagation from the RFQ to the absorber section and from just downstream of the absorber to the end of the MEBT. The RPU is a 30 mm-ID stainless-steel cylinder mounted inside a 2-3/4" CF flange. Electronics measure the amplitude of the 162.5 MHz component of the signal generated by the beam image charges and every 0.25 ms sends to the MPS the integral of the output over the last 0.25 ms window.

The reasons for choosing RPUs are their compactness, low cost, and insensitivity to the pulse length up to CW. A significant drawback is the RPU signal sensitivity to the bunch length. To overcome this, the transition to long-pulse operation always starts with cross-calibration of the RPUs. In the short-pulse mode with all scrapers in their operational positions and the MEBT kickers off, the beam is passed to the dump with low losses as measured by the difference between AC Current Transformers installed at the very beginning and end of the MEBT. Readings of all 4 RPUs in this state are taken as indication of the beam passing through with acceptable losses and constitute the baseline to which future current readings are compared to.

Typical rms noise of the differential signal was ~1%, and to eliminate corresponding trips, the threshold was typically set to 4% above the good transmission numbers defined in the procedure described above.

## LONG-PULSE OPERATION

A typical sequence for long-pulse operation begins by checking the machine with 10 µs pulses and making necessary adjustments. Then, the scrapers are moved to their operational positions, and the RPUs are cross-calibrated. After activation of all MPS channels, the pulse length is increased.

The major difficulty encountered was false trips generated by the current-comparing system, with the frequency of these trips increasing with the pulse length. The false trips were easily identified due to the fact that the RPUs were reporting a loss current exceeding the total beam current. While we were not able to find the reason for these trips or resolve them during the run, they seem to be associated with transitions between the 0.25 ms integration time windows.

The second reason for beam interruptions was real losses to the kicker protection electrodes and DPI. The threshold for such losses is set as an integral

$$Q_{th} < Q(t) = e^{-\frac{t}{\tau}} \int_0^t I_{loss}(t_1) \cdot e^{\frac{t_1}{\tau}} dt_1 \qquad (1)$$

where $I_{loss}$ is the loss current. $Q_{th}$ and $\tau$ are user-specified constants set to 50 µs·mA and 10 ms, correspondingly. By using Eq.(1), sensitivity to a given current loss increases with the pulse length. These settings allow losing the entire 5 mA beam for 10 µs pulses but limit the loss to 5 µA for CW. These interruptions were addressed by tuning the beam and adjusting the scraper positions.

Finally, any RF interruption was set to trip the beam and required manual recovery. We did not observe a dependence of the RF trips frequency on the beam pulse length. The typical behavior of some parameters during long-pulse tuning is shown in Fig. 5.

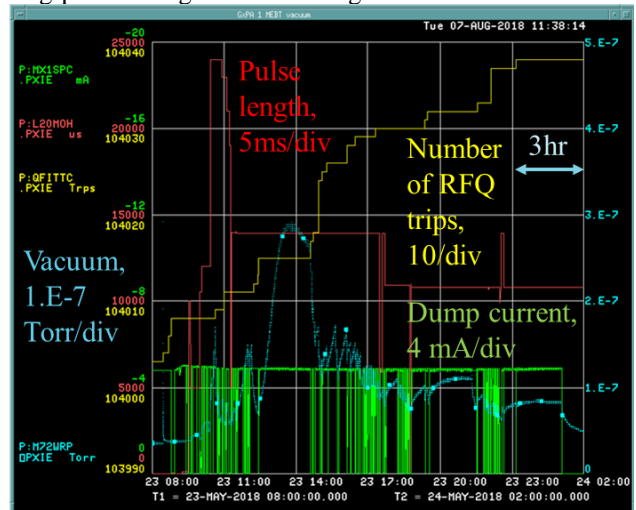

Figure 5. Operation parameters during long-pulse tuning: dump current, pulse length, RFQ trip counter, and vacuum in the most downstream bunching cavity.

The maximum pulse length reached 25 ms (50% duty factor) on the run's last day, though with < 1 min between trips. Operational stability was tested in longer runs ranging from 1 to 36 hours. Table 1 summarizes the maximum time observed between beam interruptions in some of these runs. All runs were performed with 20 Hz repetition rate.

Table 1. Parameters in Long-pulse Runs

| Pulse length, ms | Beam current, mA | Max. time between trips, hours | Configuration |
|---|---|---|---|
| 24 | 4.9 | 0.1 | 1 |
| 14 | 4.9 | 2 | 1 |
| 1.75 | 10 | 20 | 2 |
| 0.55 | 5 | 14 | 3 |

Configuration: 1- beam to the dump in the MEBT as shown in Fig. 3; 2- the entire beam is directed to the absorber prototype; 3- same as 1 but DPI is not installed and with one of the kickers pulsing ("PIP-II parameters").

## SUMMARY

Operation with high-duty factor beam at PIP2IT was extended up to 50% (5 kW of average power). For the pulse length corresponding to future PIP-II nominal parameters, 0.55 ms, the time between beam interruptions exceeded 10 hours.

## ACKNOWLEDGMENT

The authors are thankful to the many people who built the PIP2IT MEBT and helped with the long-pulse studies, including R. Andrews, C. Baffes, B. Chase, E. Cullerton, N. Eddy, J. Einstein-Curtis, M. Kucera, D. Lambert, D. Peterson, A. Saewert, J. Steimel.